\documentclass[prc,superscriptaddress]{revtex4}
\usepackage{amssymb,color,graphicx,bm}
\usepackage{amssymb}
\usepackage{amsmath}
\usepackage{graphicx}
\usepackage{hyperref}

\hypersetup{
colorlinks   = true, 
  urlcolor     = blue, 
  linkcolor    = blue, 
  citecolor   = red 
}

\definecolor{red}{rgb}{0.8,0,0}
\definecolor{RED}{rgb}{0.8,0,0}
\definecolor{violet}{rgb}{0.4,0,0.4}
\definecolor{green}{rgb}{0,0.5,0.0}
\definecolor{GREEN}{rgb}{0,0.5,0.0}
\definecolor{navy}{rgb}{0.0,0.0,0.6}
\definecolor{orange}{rgb}{0.8,0.2,0.0}
\definecolor{blue}{rgb}{0.3,0.0,0.8}

\begin{document}

\title{Revised density of magnetized nuclear matter at the neutron drip line}


\author{M. V. Vishal, Banibrata Mukhopadhyay$^*$\\
Department of Physics, Indian Institute of Science, 
Bangalore 560012, India\\ vishal.mv@physics.iisc.ernet.in , bm@physics.iisc.ernet.in\\
$^*$Corresponding Author
}

\begin{abstract}
We study the onset of neutron drip in high density matter in the
presence of magnetic field. It has been found that for
systems having only protons and electrons, in
the presence of magnetic field $\gtrsim 10^{15}$G, the neutronization occurs
at a density which is atleast an order of magnitude higher compared to that in a nonmagnetic system. In a system with heavier ions,
the effect of magnetic field, however, starts arising at a much higher
field, $\gtrsim 10^{17}$G. These results may have important implications
in high magnetic neutron stars and white dwarfs and, in general, nuclear
astrophysics when the system is embedded with high magnetic field. \\ \\


\end{abstract}

\maketitle


\section{Introduction}

The neutron stars are believed to have surface magnetic field 
as large as $\sim 10^{15}$G (magnetar model) \cite{magnetar} and hence their 
interior field ($B_{int}$) could even be a few orders of magnitude higher,
say $\sim 10^{18}$G \cite{lai,armen}, where the density is also higher. On the other hand, for white dwarfs with typical
radius ($R\sim 5000$km), the maximum $B_{int}$
could be restricted to $\sim 10^{12}$G from the scalar virial theorem.
However, for the recently proposed high density white dwarfs having highly tangled/fluctuating 
magnetic field with radius, e.g., $R\sim 70$km, $B_{int}\gtrsim 10^{17}$G \cite{prl}. 
Such smaller white dwarfs' central density ($\rho_c$) is $\sim 10^{13}$gm/cc
and hence quite above the non-magnetic threshold density of neutronization. 

All the above facts motivate us to formulate the present work and to 
study the effect of (high) magnetic field on neutronization and neutron drip for the 
degenerate fermions at high densities. This will be helpful to interpret
the detailed spectra of cooling neutron stars, radio emission and $\gamma-$ray 
burst from neutron stars. At such high density regions of neutron stars and white dwarfs,
the mean Fermi energy and the cyclotron energy of an electron exceed its rest-mass energy;
for the latter condition to hold, the magnetic field needs to be sufficiently high, making
the electrons relativistic. We know that inverse $\beta-$decay can occur when 
the energy of electrons becomes higher than the difference between the rest mass energies of
neutron and proton. Such a condition is expected to be modified in the presence of 
magnetic field which we plan to explore here at zero temperature.

Effects of magnetic field to the neutron star matter were
investigated earlier in different contexts.
For example, the enhancement of electron and neutron densities in the inner and outer 
crusts of neutron stars was discussed in the presence of high magnetic field \cite{band1,band2}.
The authors adopted the Thomas Fermi model for their calculations and
one of the motivations was to understand the transport phenomena
within magnetars.
Based on fully self-consistent
covariant density functional theory, the influence of strong magnetic 
fields on nuclear structure has been studied in the context of
magnetars \cite{dft}. These authors showed that a
field strength $\gtrsim 10^{17}$G, which presumably corresponds
to the central/crust field of magnetars, appreciably
modifies the nuclear ground state. Hence,
the composition of the magnetar crust might be radically different 
from that of normal neutron stars which might have significant 
implications to various related properties, such as
pulsar glitches, cooling etc.
Other authors investigated numerically
the variations of composition and pressure at the onset of neutron drip with the magnetic field
\cite{chem}, which we will recall in \S3 again for comparison with the present work.

The plan of the paper is the following. In the next section, we discuss 
the neutronization density in a neutron-proton-electron ($n-p-e$) gas.
Subsequently, in \S 3 we derive the neutron drip density in a system with 
heavier ions. Finally, we end with a summary in \S 4.

\section{Neutronization in a neutron-proton-electron gas}

The system is neutral in charge and, hence, the number densities of
electrons and protons are same. The generic condition for the initiation of 
inverse $\beta-$decay and hence neutronization is chemical equilibrium
given by
\begin{eqnarray} 
E^F_{e}+E^F_{p}=E^F_{n},
\label{chem}
\end{eqnarray} 
where $E^F_{e,p,n}$ are the Fermi energies for electron, proton and 
neutron respectively given by
\begin{eqnarray} 
E^F_{e,p}=\sqrt{{p^F}^2c^2+\Delta_{e,p}c^4}~~
\text{with}~~ 
\Delta_{e,p}=m_{e,p}^2(1+2\nu_{e,p}B_D^{e,p})
\end{eqnarray} 
and 
\begin{eqnarray} 
E^F_{n}=\sqrt{{p^F_n}^2c^2+m_n^2c^4},
\end{eqnarray} 
where $m_{e,p,n}$ are the masses of electron, proton and neutron respectively,
$B_D^{e,p}$ are the magnetic fields in the units of respective critical fields 
$B_c^e=4.414\times 10^{13}$G and $B_c^p=1.364\times 10^{20}$G, 
for electrons and protons respectively, $\nu_{e,p}$ are the occupied Landau 
levels in the respective energies, $p^F$ is the Fermi momentum of electrons and protons,
$p^F_n$ is the Fermi momentum of neutrons and $c$ is the speed of light.
The neutronization density for this system was investigated by previous 
authors \cite{lai}. Nevertheless, here we discuss the results categorically.
Let us consider three possible cases, some of them may turn out to be absurd.

\paragraph{Case I: ${p^F}^2/c^2>>\Delta_{e,p}$:} This happens when both
protons and electrons are highly relativistic. At the beginning of formation of neutrons (and hence the corresponding momentum
 is zero),
from eqn. (\ref{chem}) we obtain 
\begin{eqnarray} 
p^F/c= m_n/2,
\label{rel2}
\end{eqnarray}
which however violates initial choice as $m_p\sim m_n$. 
Hence, this situation is unphysical to reveal the neutronization.

\paragraph{Case II: ${p^F}^2/c^2\sim\Delta_p>>\Delta_{e}$:} 
This happens when 
electrons are highly relativistic but protons are just becoming 
relativistic. From eqn. (\ref{chem}) the initiation of neutronization occurs at
\begin{eqnarray} 
p^F/c= \frac{m_n^2-\Delta_p}{2m_n},
\label{rel2}
\end{eqnarray} 
which, however, turns out to be negative unless $\Delta_p\sim m_p^2$.
However, this corresponds to $E^F_{p}\sim m_pc^2$ which cannot reveal
relativistic protons. Hence, this case is unphysical.

\paragraph{Case III: ${p^F}^2/c^2\sim\Delta_e<<\Delta_{p}$:} 
This corresponds to relativistic electrons but nonrelativistic protons,
which reads eqn. (\ref{chem}) as 
\begin{eqnarray} 
\Delta_p^{1/2}+\sqrt{\left(\frac{p^{F}}{c}\right)^2+\Delta_e}=m_n
\label{relchem1}
\end{eqnarray} 
at the neutronization, revealing further
\begin{eqnarray} 
p^F/c= m_n\sqrt{\left(1-\frac{m_p}{m_n}\sqrt{1+2\nu_pB_D^p}\right)^2-\left(\frac{m_e}{m_n}
\right)^2(1+2\nu_eB_D^e)}.
\label{rel1}
\end{eqnarray} 
When the Fermi energy and magnetic field of the system are such that 
the electrons lie in the ground Landau level only (with $\nu_e=0$)
and the protons are not affected by the magnetic field (with $E_{Fp_{max}}\sim m_pc^2$), the neutronization density 
\begin{eqnarray} 
\rho_{neut}=\frac{2eB_D^eB_c^e}{h^2c}p^F(m_e+m_p)=0.343B_D^e10^{7}~\text{gm/cc},
\label{rel11}
\end{eqnarray} 
where $e$ is the charge of electrons and $h$ the Planck's constant.
Eq. (\ref{rel11}) implies that the neutronization density increases linearly
with the magnetic field and for the increment to be at least an order of magnitude
compared to the nonmagnetic case, the magnetic field $B$ has to be $\sim 10^{15}$G.

However, beyond neutronization, the chemical equilibrium eqn. (\ref{chem}) gives
\begin{eqnarray}
\sqrt{{p^F}^2+m^2_ec^2}+\sqrt{{p^F}^2+m^2_pc^2}&=&\sqrt{{p^F_n}^2+m^2_nc^2}\label{bd},
\end{eqnarray}
where we have assumed that all the electrons reside in the ground Landau level and 
for this to be true, $B_D^e>2.77$ (see \cite{lai}), and the protons are not
affected by the chosen value of magnetic field.\\

\begin{figure*}
\includegraphics[scale=0.5,angle=-90]{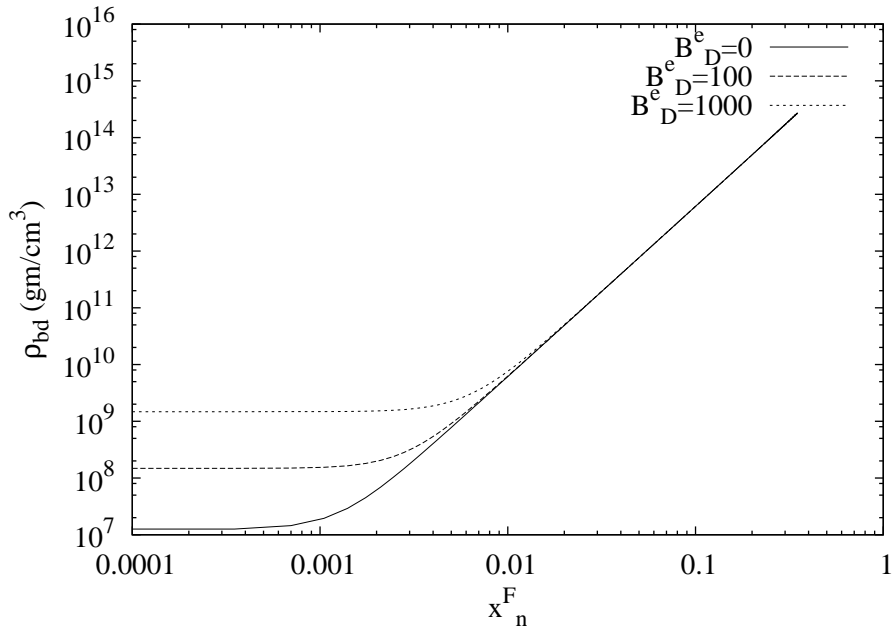}
\caption{
Density beyond neutronization in an $n-p-e$ system 
as a function of (dimensionless) Fermi momentum of neutrons, for
\noindent different magnetic fields. All electrons reside in the ground Landau level
and the protons are not affected by the chosen magnetic fields.
}
\label{beyd}
\end{figure*}

Therefore, the density beyond neutronization is given by
\begin{eqnarray}
\rho_{bd}\approx  m_pN_e+\frac{\epsilon_n}{c^2},~~~
\text{where~~~~~}N_e=\frac{2B_D^e}{(2\pi)^2\lambda_e^3}\,\frac{p^F}{m_ec},
\end{eqnarray}
where
\begin{eqnarray}
\nonumber
\epsilon_n &=& \frac{m_nc^2}{\lambda_n^3}\chi(x^F_n),~~x^F_n = p^F_n/m_nc,
~~\lambda_e=\hbar/m_ec,~~\lambda_n=\hbar/m_nc,\\
\nonumber
\chi(x^F_n) &=&\frac{1}{8\pi^2}\left[x^F_n(1+2{x^F_n}^2)\sqrt{1+{x^F_n}^2}-\log(x^F_n+\sqrt{1+
{x^F_n}^2}) \right].\\
\end{eqnarray}
Here $N_e$ is the number density of electrons (which is also the number 
density of protons), $\epsilon_n$ is the energy density of neutrons,
$\lambda_e$ and $\lambda_n$ are the Compton wavelengths of the electron and
neutron respectively and $\hbar=h/2\pi$. Now eqn.
\eqref{bd} expresses $p^F$ in terms of $p^F_n$. Hence $\rho_{bd}$ varies 
as a function of $x^F_n$ for a given $B_D^e$, as shown in Fig. \ref{beyd}.


\section{Neutron drip in a gas with heavy ions}

The results of this section are mostly applicable for neutron stars and maybe
for white dwarfs with heavier ion(s).
We adopt the Harrison-Wheeler formalism \cite{HW} in describing the equilibrium
composition of nuclear matter. We do not include the shell effects in the model, as they have 
a negligible effect on the density of the system. Here we intend to determine the variation in the
density at the onset of drip with the variation in magnetic field. Hence, the lattice energy 
effects can be neglected, as they have a significant influence only on the composition of the system and an
even lesser influence on the pressure. They do not alter the density of the mixture \cite{st}. 
Moreover, the Harrison-Wheeler model has the added advantage of
allowing one to obtain a purely analytical relation between the drip density and the field strength, as we shall show
below. Previous authors obtained a relation between the field and
drip pressure (see, e.g., \cite{chem}), however purely in numerical computations. 
Their results have the disadvantage that the drip pressure versus magnetic
field relation is obtained numerically at discreet values of the field
with reasonably large intervals.
This kills an important physics revealing oscillations of the drip density
at the lower values of field, which we address below in detail in our analysis.
Moreover, they considered only a limited
range of fields while studying the drip pressure variation. Whereas we investigate the density for a wide range of 
field strengths so as to understand the regimes where there is a significant change in
the drip density.

We assume that given enough time, following nuclear burning, cold catalyzed material will achieve complete 
thermodynamic equilibrium. The matter composition and equation of state will then be
determined by the lowest possible energy state of the matter. The constituents of the matter are assumed to 
be ions, free electrons and free neutrons and the drip density is obtained by
finding the density of the equilibrium composition, when the neutron number density is zero (neutrons just 
start appearing). We further assume the strong magnetic field present in
the matter is constant throughout. We then obtain the equilibrium density by minimizing the energy density 
of the mixture with respect to the number densities of the ions, electrons and
neutrons. Due to the presence of the magnetic field, the electrons are Landau quantized but the field 
of present interest (say, upto the value which a neutron star core could plausibly have) is small enough 
so that it does not affect the
ions. 

Now we proceed to establish our results. Note that for the convenience of the readers, 
we follow the well-known textbook by Shapiro \& Teukolsky \cite{st} in
developing our results, which discusses the non-magnetic results in
detail.
The total energy density of the system can be written as
\begin{equation}
\epsilon_{total} = N_{ion} M(A,Z)+\epsilon_e^{'}(N_e)+\epsilon_n(N_{n}) \label{energy}.
\end{equation}
Here $M(A,Z)$ is the energy of a single ion $(A,Z)$, including the rest mass energy, where $A$ and $Z$ 
are atomic mass and atomic number respectively.
It is conventional to include the rest mass energy of the electrons in the energy of 
the ions (see, e.g., \cite{st} for details of non-magnetic results). Therefore, $\epsilon_e^{'}$ is the energy density of the electrons after subtracting $N_em_ec^2$
from the total energy density $\epsilon_e$ and 
 $\epsilon_n$ is the total energy density of neutrons. The system is neutral
in charge. Hence,
the number densities $N_{ion}$, $N_e$ and $N_{n}$, of ions, electrons and neutrons 
respectively, and the total baryon number density $N$ are related as 
\begin{eqnarray}
N=N_{ion}A+N_{n},\,\,
N_e=N_{ion}Z.
\end{eqnarray}
These can be written in terms of mean numbers per baryon as
\begin{eqnarray}
Y_{ion}A+Y_{n}=&1,\,\,
Y_{ion}Z=Y_e, \label{composition}
\end{eqnarray}
where $Y_i=N_i/N$. 
Hence at $T=0$, we can write $\epsilon_{total}$ as a function of 
either $(N,A,Z,Y_n)$ or $(N,Y_{ion},Y_e,Y_n)$.
The equilibrium composition is obtained by minimizing $\epsilon_{total}$ with respect to $A$, $Z$ and $Y_n$ for a constant $N$.

We use the semi-empirical mass formula of Green \cite{green} based on the liquid drop nuclear model
\begin{equation}
M(A,Z)=m_uc^2 \left[b_1A+b_2A^{2/3}-b_3Z+\frac{b_4}{A}\left(\frac{A}{2}-Z\right)^2+\frac{b_5Z^2}{A^{1/3}}\right],
\end{equation}
where
\begin{equation}
 b_1=0.991749\text{,~~} b_2=0.01911\text{,~~} b_3=0.000840 \text{,~~}
 b_4=0.10175\text{,~~} b_5=0.000763 \nonumber
\end{equation}
and $m_u=1.66 \times 10^{-24}$gm (1 amu).
Hence eqn. \eqref{energy} becomes
\begin{equation}
\epsilon_{total} = \frac{N(1-Y_n)}{A} M(A,Z)+\epsilon_e^{'}(N_e)+\epsilon_n(N_{n}).
\label{epstot}
\end{equation}
Moreover,
\begin{eqnarray}
\dfrac{\mathrm{d}\epsilon^{'}_e}{\mathrm{d}N_e}=E^F_e-m_ec^2 \label{efe},\,\,\,
\dfrac{\mathrm{d}\epsilon_n}{\mathrm{d}N_{n}}=E^F_n.
\label{depsdn}
\end{eqnarray}
Now we approximate $A$ and $Z$ as continuous variables and equations for 
equilibrium compositions can be found from eqns. \eqref{epstot} and \eqref{depsdn} as
\begin{eqnarray}
\frac{\partial\epsilon_{total}}{\partial Z}=0,\,\,\,{\rm and\,\,hence}\,\,\, 
\frac{\partial M}{\partial Z}=-(E^F_e-m_ec^2), \label{MwithZ}
\end{eqnarray}
\begin{eqnarray}
\frac{\partial\epsilon_{total}}{\partial A}=0,\,\,\,{\rm and\,\,hence}\,\,\, 
A^2\frac{\partial}{\partial A}\left(\frac{M}{A}\right)=Z\left(E^F_e-m_ec^2\right),
\label{a2d}
\end{eqnarray}
and finally
\begin{eqnarray}
\frac{\partial\epsilon_{total}}{\partial Y_n}=0 ,\,\,\,\text{and hence using eqn. \eqref{a2d}}\,\,\,
\frac{\partial M}{\partial A}=E^F_n.
\end{eqnarray}

\begin{figure}
\includegraphics[scale=0.5,angle=-90]{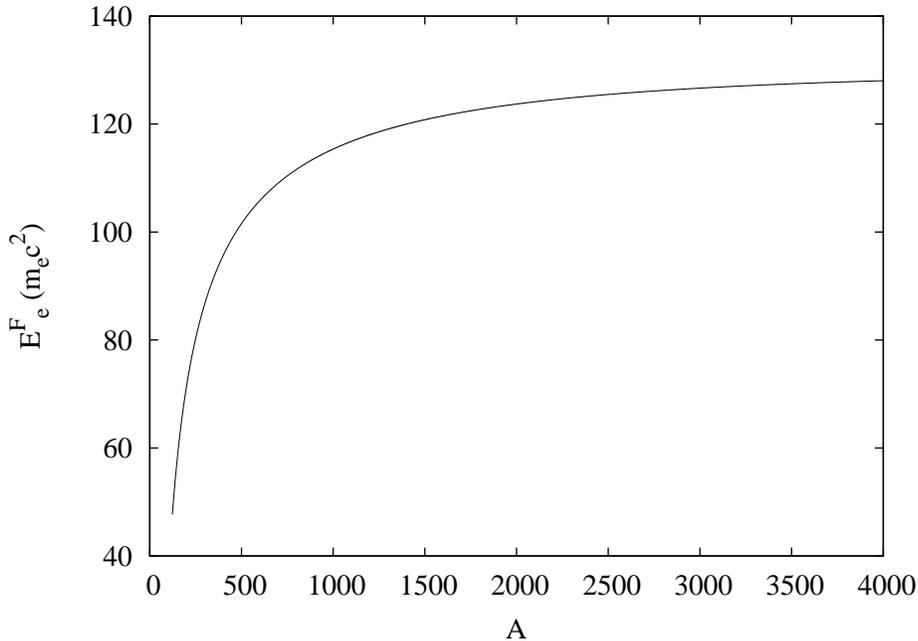}
\caption{Electron Fermi energy beyond drip as a function of atomic mass.}
\label{fermi-A}
\end{figure}

Now combining eqns. \eqref{MwithZ} and \eqref{a2d}, we obtain
\begin{eqnarray}
Z=\left(\frac{b_2}{2b_5}\right)A^{1/2}. \label{equation1}
\end{eqnarray}
Eqns. \eqref{MwithZ} and \eqref{a2d} also give
\begin{eqnarray}
\left[b_3+b_4\left(1-\frac{2Z}{A}\right)-2b_5\frac{Z}{A^{1/3}}\right]m_uc^2&=&E^F_e-m_ec^2\label{electron},\label{elec}
\end{eqnarray}
\begin{eqnarray}
\left[b_1+\frac{2}{3}\,b_2\,A^{-1/3}+b_4\left(\frac{1}{4}-\frac{Z^2}{A^2}\right)-\frac{b_5Z^2}{3A^{4/3}}\right]m_uc^2&=&E^F_n, \label{neutron}
\end{eqnarray}
where
$E^F_n=\sqrt(1+{x^F_n}^2)m_nc^2$, and at the onset of neutron drip, $x_n=0$ when the number density of neutron is just zero. 
Therefore, at the onset of drip, eqn. \eqref{neutron} can be written as
\begin{equation}
b_1+\frac{2}{3}\,b_2\,A^{-1/3}+b_4\left(\frac{1}{4}-\frac{Z^2}{A^2}\right)-\frac{b_5Z^2}{3A^{4/3}}=\frac{m_n}{m_u}. \label{equation2}
\end{equation}
Solving eqns. \eqref{equation1} and \eqref{equation2}, we obtain $(A,Z)\approx(122,39.1)$ for the onset of drip. Using these values for $A$ and $Z$, 
we can obtain the electron Fermi energy at the drip from eqn. \eqref{electron} as
\begin{eqnarray}
\frac{E^F_{ed}}{m_ec^2}&=&47.2367.\label{equation3}
\end{eqnarray}
 Figure \ref{fermi-A} shows, combining eqns.
(\ref{equation1}) and (\ref{elec}), that beyond drip, $E^F_e$ first
increases with the increase of $A$ and subsequently saturates to a value
$\sim 128.3$ at a large $A$. However, a large $A$ at a high magnetic field
corresponds to a value of $E^F_e$ and then density well within the 
maximum possible central density of a neutron star. For example, at
$B_D^e=20000$ (and hence the magnetic field $8.83\times 10^{17}$G), 
$A=122$ corresponds to a density $\rho_{bd}=4.39\times 10^{12}$gm/cc,
whereas the corresponding nonmagnetic $\rho_{bd}=3.25\times 10^{11}$gm/cc.
This clearly indicates the huge effects of magnetic field in the matter.
Higher values of $A=600$ and $4000$, at the same magnetic field, correspond to 
$\rho_{bd}=1.2\times 10^{13}$gm/cc and $7\times 10^{13}$gm/cc respectively ---
all are within the typical nuclear matter density $\sim 10^{14}$gm/cc.
Hence, highly magnetized neutron stars may exhibit atomic structures with large $A$.

The Fermi energy of a Landau quantized electron with magnetic field strength 
$B$ confined in $\nu_e$ levels can be rewritten as
\begin{eqnarray}
E^F_e &=&\left[c^2p^F_{ze}(\nu_e)^2+m_e^2c^4\left(1+\frac{2\nu_e B}{B^e_c}\right)\right]^{1/2},
\nonumber
\end{eqnarray}
where $p^F_{ze}$ is $z$-component of the Fermi momentum of the electron.
Therefore, from eqn. \eqref{equation3}
\begin{eqnarray}
x^F_e(\nu_e)=\frac{p^F_{ze}}{m_ec}=\left(2230.31-\frac{2\nu_e B}{B^e_c}\right)^{1/2}. \label{xe}
\end{eqnarray}

Let us now determine the number of states occupied by the Landau quantized electrons \cite{lai}.
This is given by
\begin{eqnarray}
N_e=\int dN_e= \sum_{\nu_e=0}^{\nu_m} \frac{eB}{h^2c}\,g_{\nu_e}\int dp_z
=\frac{2B^e_D}{(2\pi)^2\lambda_e^3}\sum_{\nu_e=0}^{\nu_m} g_{\nu_e}\, x^F_e(\nu_e),
\label{nee}
\end{eqnarray}
where 
$g_{\nu_e}$ 
the degeneracy factor which is $2-\delta_{\nu_e,0}$. The upper limit $\nu_m$ in the summation is obtained from the condition that ${p^F_{ze}(\nu_e)}^2\geq 0$, which gives
\begin{eqnarray}
\nu_e \leq  \frac{(E^F_e/m_ec^2)^2-1}{2B^e_D}\,\,\,\text{ and hence}\,\,\,
\nu_m = \frac{(E^F_{emax}/m_ec^2)^2-1}{2B^e_D}.
\end{eqnarray}

\begin{figure}
\vspace{-30pt}
\includegraphics[scale=0.5,angle=-90]{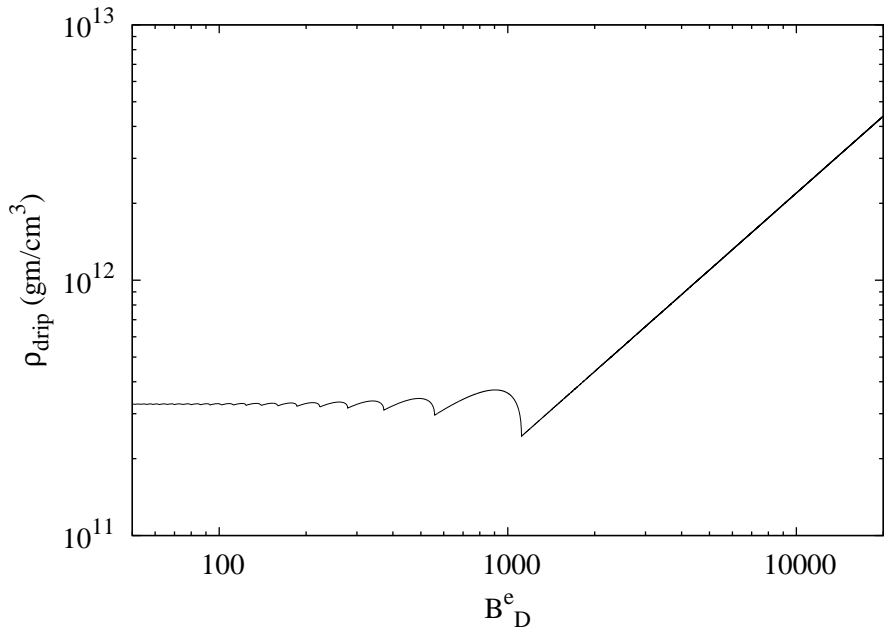}
\caption{Drip density in a gas of electrons and ions as a function of magnetic field.}
\label{ionfig}
\end{figure}

The electron energy density at zero temperature is then \cite{lai}
\begin{eqnarray}
\epsilon_e &=& \frac{2B_D^e}{(2\pi)^2\lambda_e^3}\sum_{\nu_e=0}^{\nu_m} g_{\nu_e}\int_0^{x^F_e(\nu_e)} E^F_e \, \mathrm{d}\left(\frac{p_{ze}}{m_ec}\right)\nonumber\\
&=& m_ec^2\frac{2B^e_D}{(2\pi)^2\lambda_e^3} \sum_{\nu_e=0}^{\nu_m} g_{\nu_e} (1+2\nu_e B^e_D)\,\Psi\left(\frac{x^F_e(\nu_e)}{(1+2\nu_e B^e_D)^{1/2}}\right),
\end{eqnarray}
where
\begin{equation}
\Psi(z)=\frac{1}{2}z\sqrt{1+z^2}+\frac{1}{2}\,ln(z+\sqrt{1+z^2}).\nonumber
\end{equation}
The density of the mixture of ions and electrons at the onset of drip is therefore
\begin{eqnarray}
\rho_{drip}=\frac{\epsilon_{total}}{c^2}
=\frac{N_e M(A,Z)/Z+\epsilon_e-N_em_ec^2}{c^2}.
\end{eqnarray}
Now eqn. \eqref{xe} gives $x_e^F$ as a function of $B^e_D$. Therefore, the above equation gives us a relationship between $\rho_{drip}$ and $B^e_D$. Note that for $B>>B_c^e$, $\nu_e=0$ and hence $x_e^F$ becomes independent of $B$, which
renders $N_e$ to be linear in $B^e_D$ in eqn. (\ref{nee}). Indeed, Fig. 
\ref{ionfig} shows that for $B\gtrsim 10^{16}$G, drip density 
increases linearly with magnetic field.
This linearity arises 
due to the fact that beyond a certain $B^e_D$, the Landau quantized 
electrons reside in the ground state energy level. We also notice that 
before the onset of linearity, the drip density oscillates about the 
non-magnetic result, with small amplitudes. 
A significant change in the drip density arises only in the linear regime. 
For instance, the drip density increases by an 
order, compared to that of the nonmagnetic gas, for a field 
$\sim 6.4\times 10^{17}$G. 

\begin{figure}
\vspace{20pt}
\includegraphics[scale=0.5,angle=-90]{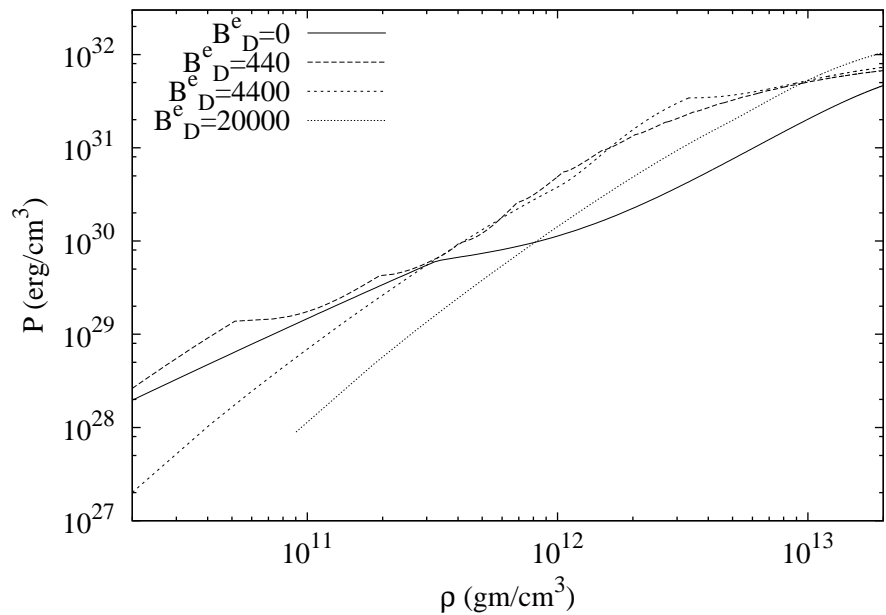}
\caption{Equation of state for $B_D^e=440, 4400, 20000$ and nonmagnetic
matter.}
\label{eos}
\end{figure}

For completeness, we also provide in Fig. \ref{eos} the equation of states 
(EoSs) of magnetized 
nuclear matter for different values of (fixed) magnetic field
and compare them with the nonmagnetic EoS. All the EoSs are plotted starting from
the density corresponding to $A=56$. However, for a very high $B$ 
(here $B_D^e=20000$), the corresponding density increases noticeably 
compared to that of a lower $B$ and hence the corresponding EoS begins
from a larger density, as shown in Fig. \ref{eos}.

\section{Summary}

We have studied the effects of magnetic field on the onsets of neutronization
for $n-p-e$ and neutron drip for ion-electron gas systems. We have found that beyond a certain 
value of magnetic field, the neutronization/drip density of either of the systems increases, compared to the nonmagnetic cases,
linearly with the magnetic field. This value of field for the former is 
$B^e_D \sim 2.7$ and for the latter $\sim 1115$. 
The significant change in the drip density arises only in the linear regime. 
For instance, the drip density for the ionic system increases by an 
order, compared to that of the nonmagnetic gas, for a field 
$\sim 6.4\times 10^{17}$G. 

Apart from the various applications to neutron stars, the present 
findings may have interesting consequences to 
the recently proposed high density ($\gtrsim 10^{11}$gm/cc), high magnetic field ($\gtrsim 10^{15}$G) 
white dwarfs \cite{prl,prd,apj,mpla,jcap}. The inner region of white dwarfs, in 
particular for $B_{int}\gtrsim 10^{17}$G, would have been neutronized if the drip density had not been changed with the field.
Note that some authors have questioned the existence of such white dwarfs \cite{chem2}.
For example, the authors have argued that
for the white dwarfs to be stable against neutronization, the magnetic field at the center should be 
less than few times $10^{16}$G for typical matter compositions. However, as discussed in detail
in \cite{mpla} (also see \cite{prd}), for the central field $\sim 10^{16}$G, the mass
of the said highly magnetized white dwarfs already becomes significantly
super-Chandrasekhar with the value $2.44$ solar mass. Therefore, such super-Chandrasekhar
white dwarfs remain stable according to the neutronization limit given by the same authors only.
Even if the white dwarfs with very large field consist of central region
with density larger than the modified drip density, the present 
finding will help in putting a constraint on such models of white dwarfs
with pure electron degenerate matter.

\section*{Acknowledgment} 
The work has been supported by the project with Grant No. ISRO/RES/2/367/10-11.
The authors would like to thank the referee for useful comments and 
suggestions to improve the quality of the paper.


\end{document}